\documentclass[apjl]{emulateapj}
\usepackage{apjfonts}

\usepackage{amsmath}
\usepackage{amssymb}
\usepackage{epsfig}
\usepackage{natbib}

\newcommand{\kms}{\,{\rm km\,s^{-1}}}
\newcommand{\au}{\,{\rm AU}}

\newcommand{\yr}{\,{\rm yr}}

\newcommand{\cm}{\,{\rm cm}}

\newcommand{\pc}{\,{\rm pc}}
\newcommand{\kpc}{\,{\rm kpc}}

\newcommand{\s}{\,{\rm s}}

\newcommand{\jy}{\,{\rm Jy}}
\newcommand{\mjy}{\,{\rm mJy}}
\newcommand{\mujy}{\,\mu{\rm Jy}}
\newcommand{\ghz}{\,{\rm GHz}}

\newcommand{\msun}{\,M_\odot}

\newcommand{\lsun}{\,L_{\rm \odot}}

\newcommand{\be}{\begin{equation}}
\newcommand{\ee}{\end{equation}}

\newcommand{\bea}{\begin{eqnarray}}
\newcommand{\eea}{\end{eqnarray}}

\renewcommand{\arcsec}{^{\prime\prime}}

\renewcommand{\farcs}{.\!\!^{\prime\prime}}
\renewcommand{\fdg}{.\!\!^{\circ}}

\newcommand{\ben}{\begin{enumerate}}
\newcommand{\een}{\end{enumerate}}

\newcommand{\calm}{M_{6.5}}    
\newcommand{\calpo}{P_1}   

\begin{document}

\shorttitle{Radio Pulsars Orbiting Sgr A$^*$}
\shortauthors{PFAHL \& LOEB}


\submitted{Submitted to The Astrophysical Journal}

\title{Probing the Spacetime Around Sgr A$^*$ with Radio Pulsars}

\author{Eric Pfahl\altaffilmark{1} and Abraham Loeb\altaffilmark{2}}

\affil{Harvard-Smithsonian Center for
Astrophysics, 60 Garden Street, Cambridge, MA 02138; \\
epfahl@cfa.harvard.edu, aloeb@cfa.harvard.edu}

\altaffiltext{1}{Chandra Fellow}
\altaffiltext{2}{Guggenheim Fellow}


\begin{abstract}

The supermassive black hole at the Galactic center harbors a bound
cluster of massive stars that should leave neutron-star remnants.
Extrapolating from the available data, we estimate that $\sim$1000
radio pulsars may presently orbit Sgr A$^*$ with periods of
$\la$100\,yr.  Optimistically, 1--10 of the most luminous of these
pulsars may be detectable with current telescopes in periodicity
searches at frequencies near 10\,GHz, where the effects of
interstellar scattering are alleviated.  Long-term timing observations
of such a pulsar would clearly reveal its Keplerian motion, and
possibly show the effects of relativistic gravity.  We briefly discuss
how pulsar timing can be used to study the dynamical and interstellar
environment of the central black hole, and speculate on the prospects
for astrometric observations of an orbiting pulsar.

\end{abstract}




\section{INTRODUCTION}\label{sec:intro}

Ten years of near-infrared observations of the Galactic center have
revealed the proper motions of nearly two dozen stars within $0\farcs
5$ of the compact radio source Sgr A$^*$.  Significant astrometric
accelerations measured for 8 members of the Sgr A$^*$ stellar cluster
point to a common center of gravity coincident with the position of
Sgr A$^*$, and imply a central mass of (3--$4)\times 10^6\msun$
\citep{ghez03b,schodel03}.  Stars S0-2 and S0-16 have the most compact
orbits yet identified, with respective periods of $\simeq$15\,yr and
$\simeq$30\,yr, eccentricities of $\simeq$0.88 and $\simeq$0.95, and
comparable pericenter distances of $\simeq$100\,AU
\citep{schodel02,ghez03b,eisenhauer03}.  If the central mass is
confined within 100\,AU, the implied density is
$\ga$$10^{16}\msun\pc^{-3}$, which essentially rules out existing
models alternative to the hypothesis that the central object is a
supermassive black hole \citep[BH;][]{maoz98,ghez03b,schodel03}.

Evidence from the near-infrared spectrum of S0-2 \citep{ghez03a} and
the integrated spectrum within $\simeq$$0\farcs 5$ of Sgr A$^*$
\citep{genzel97,eckart99,figer00,gezari02} suggest that the observed
Sgr A$^*$ stellar cluster is largely comprised of luminous
($\sim$$10^4\lsun$), early-type (O9 to B0) stars.  If these stars are
near the main sequence, their masses are 10--$20\msun$. How these
stars came to reside so near the supermassive BH remains a puzzle; for
discussions and references, see \citet{genzel03} and \citet{ghez03b}.
Nevertheless, the existence of a cluster of massive stars tightly
bound to Sgr A$^*$ has important implications.

Stars of mass 10--$20\msun$ have nuclear lifetimes of $\sim$$10^7\yr$,
and leave neutron-star (NS) remnants.  Therefore, we expect a
significant number of NSs to be bound to Sgr A$^*$ in orbits similar
to those of the observed cluster stars, as well as in more compact
orbits.  Source confusion so far inhibits the discovery of stars with
orbital periods of $\la$$10\yr$ about Sgr A$^*$
\citep{genzel03,ghez03b}, though we anticipate that massive stars and
NSs populate this region.  The most exciting possibility is if some of
the NSs orbiting Sgr A$^*$ are detectable radio pulsars, an idea first
considered in the prescient article by \citet{paczynski79}.  In
\S~\ref{sec:pop}, we estimate the total number of {\em normal} radio
pulsars---i.e., those with surface magnetic field strengths of
$\sim$$10^{11}$--$10^{13}$\,G---that may presently orbit the central
BH with periods of $\la$100\,yr.

Radio-wave scattering in the interstellar plasma poses the largest
obstacle to discovering pulsars near Sgr A$^*$, where column the
density of free electrons is very high.  At observing frequencies of
$\simeq$1\,GHz, pulsed emission from this vicinity suffers severe
temporal broadening, prohibiting the detection of pulsars as periodic
sources \citep{cordes97}.  Relatively high frequencies of $\ga$10\,GHz
are required to alleviate the effects of scattering.  These issues are
addressed in \S~\ref{sec:det}, where we estimate the number of pulsars
orbiting Sgr A$^*$ that may be detectable with current telescopes.

The Keplerian motion of a pulsar orbiting Sgr A$^*$ would be clearly
apparent in its long-term timing properties.  Relativistic gravity may
introduce measurable deviations from the best-fit Keplerian timing
solution, depending on the orbital parameters and timing precision.
Various arrival-time delays and secular effects are quantified in
\S~\ref{sec:timing}.  In \S~\ref{sec:dis}, we consider what pulsar
timing can teach us about the accretion flow onto the Galactic BH and
the stellar dynamical environment, and investigate the possibility of
astrometrically monitoring an orbiting pulsar.


\section{RADIO PULSARS ORBITING SGR A$^*$}\label{sec:pop}

The observed Sgr A$^*$ stellar cluster occupies the central
$\simeq$4000\,AU about the BH.  Although the observational census is
incomplete, this volume likely contains at least several dozen massive
stars.  However, it is not the present population of massive stars,
but rather their predecessors, that are the progenitors of radio
pulsars orbiting Sgr A$^*$.  Since the origin of the observed cluster
stars is unknown, we can only speculate on the history of the
population of massive stars in this region.  It is plausible that
cluster stars are steadily or episodically replenished as they evolve
and leave NS remnants.  This might be the case if the mechanism that
feeds stars into the central $\simeq$4000\,AU is linked to the
significant star-formation activity on larger scales.  Observational
evidence suggests that the central $\sim$200\,pc of the Galaxy is a
region of past and current star formation, with an average
massive-star formation rate as large as $\sim$$10^{-3}\yr^{-1}$, or
$\sim$10\% of the rate in the entire Galaxy (Mezger et al. 1999;
Launhardt, Zylka, \& Mezger 2002).

A plausible initial guess is that there are roughly as many pulsars
within $0\farcs 5$ of Sgr A$^*$ as there are massive stars.  Thus,
perhaps several tens of radio pulsars, with ages of $\la$$10^7\yr$,
orbit near the black hole.  A simple extension of this argument
provides an estimate of the total number of active radio pulsars
orbiting Sgr A$^*$, including those much too faint to be detected by
current telescopes.  Assume that over a time of $\ga$$10^8\yr$ an
average number of $\sim$10--100 NS progenitors orbit Sgr A$^*$ with
semimajor axes of $\la$4000\,AU and periods of $P_{\rm orb} \la
100\yr$.  A stellar lifetime of $\sim$$10^7\yr$ then implies a NS
birthrate of $\sim$$10^{-6}$--$10^{-5}\yr^{-1}$.  We further suppose
that a large fraction of NSs turn on as radio pulsars shortly after
birth.  Radio emission terminates when a pulsar crosses the ``death
line'' in the $\log P_p$--$\log \dot{P}_p$ plane, where $P_p$ is the
pulse period (e.g., Rudak \& Ritter 1994, and references therein).  A
coarse inspection of pulsar statistics in the $\log P_p$--$\log
\dot{P}_p$ plane shows that a median terminal age for normal radio
pulsars is probably $\sim$$10^8\yr$, depending on the model for the
death line, and neglecting magnetic-field decay
\citep[e.g.,][]{bhat92,rudak94,tauris01}.  Therefore, we predict that
$\sim$100--1000 radio pulsars presently orbit Sgr A$^*$ with $P_{\rm
orb} \la 100\yr$ and ages of $\la$$10^8\yr$.

Typical NS ``kick'' speeds of $v_k \simeq 100$--$300\kms$ are inferred
from the proper motions of $\sim$100 isolated pulsars in the Galactic
disk \citep[e.g.,][]{hansen97,arzoumanian02}. Since the stars we are
considering have orbital speeds of $v_{\rm orb} \ga 1000\kms$, an
impulsive NS kick will usually cause only a small fractional change
($\sim$$v_k/v_{\rm orb}$) in the orbital parameters.  Therefore, the
distributions of pulsar orbital parameters should be similar to those
of their progenitors.  For a stellar number density $n(r)\propto
r^{-3(1 + q)/2}$ about Sgr A$^*$, the differential period distribution
is $p(P_{\rm orb}) \propto P_{\rm orb}^{-q}$ if the velocity
distribution is isotropic \citep[e.g.,][]{schodel03}.
\citet{genzel03} find that $q \simeq 0$ at $\la$$10\arcsec$ from Sgr
A$^*$, so that $p(P_{\rm orb})$ is approximately flat.  If there are
$\sim$1000 radio pulsars with $P_{\rm orb}$ uniformly distributed
over, e.g., 1--100\,yr, then $\sim$100 pulsars may have $P_{\rm orb}
\la 10\yr$.


\section{PULSAR DETECTION}\label{sec:det}

Scattering of radio waves by fluctuations in the free-electron density
causes angular broadening of radio images and temporal smearing of
pulsed emission.  Toward the Galactic center, observed scattering
diameters of $\simeq$$1\arcsec\,(\nu/{\rm GHz})^{-2}$ for Sgr A$^*$
and nearby OH masers imply a scattering timescale of
$\sim$$300\s\,(\nu/{\rm GHz})^{-4}$ \citep{lazio98a,lazio98b}, where
$\nu$ is the observing frequency.  Pulsars with $P_p \la 1\s$ would
thus be undetectable as periodic sources for $\nu \simeq 1\ghz$.  At
$\nu \ga 5\ghz$, the scattering time is $\la$$P_p$, but the flux
density is reduced according to the declining power-law spectrum,
$S_\nu \propto \nu^{-\alpha}$, followed by most pulsars.  Cordes \&
Lazio (1997; see also Kramer et al. 2000) find that near Sgr A$^*$ the
pulsed flux is maximized at $\nu \simeq 10\ghz$ for $P_p \simeq 1\s$,
with a weak dependence on $P_p$ and $\alpha$.  At higher frequencies,
the pulsed fraction of the flux is $\simeq$1, and the pulse duty
cycle, $\epsilon$, approaches its intrinsic value; $\epsilon \simeq
0.05$ is typical for normal pulsars.

The minimum detectable flux density is $S_{\rm min} \simeq C\,S_n
[\epsilon/(N_p\Delta\nu\,t_{\rm int})]^{1/2}$
\citep[e.g.,][]{dewey85}, where $C \simeq 10$ is the signal-to-noise
threshold, $S_n$ is the noise flux from the telescope and sky, $\Delta
\nu$ is the bandwidth, $N_p = 2$ is the number of polarizations, and
$t_{\rm int}$ is the integration time.  For the 100-m Robert C. Byrd
Green Bank Telescope\footnote{\url{http://www.gb.nrao.edu/GBT/}}
operating at $\nu \ga 10\ghz$, the noise flux is $S_n \simeq
25$--$50\jy$ toward the Galactic center, and $\Delta\nu \simeq 1\ghz$.
We then find that 
\begin{equation}
S_{\rm min} \simeq
(20-40)\mujy\,\left(\frac{\epsilon}{0.05}\right)^{1/2}
\left(\frac{t_{\rm int}}{1\,{\rm hr}}\right)^{-1/2}~.
\end{equation}
From Green Bank, the Galactic center can be observed for up to
$\simeq$8\,hr per day (S. Ransom, private communication), so that
sensitivities of $S_{\rm min} \simeq 10\mujy$ may be possible.

Observed Galactic pulsars have intrinsic 400-MHz luminosities of
$(1$--$10^4)\mjy\kpc^2$, with a cumulative distribution
$f(>\!\!L_{400}) \simeq L_{400}^{-1}$
\citep[e.g.,][]{lyne85,taylor93,lyne98}, where $L_{400}$ is in
mJy\,kpc$^2$ and is assumed to be $\ll$$10^4$.  Pulsar luminosities
are typically defined by $L_\nu = D^2S_\nu$ \citep[e.g.,][]{taylor77},
where $D$ is the distance.  For a pulsar with $S_\nu \propto
\nu^{-\alpha}$ at the 8-kpc distance of Sgr A$^*$
\citep{eisenhauer03}, we have $L_{400} \simeq
64\,S_\nu\,x^{\alpha}\mjy\kpc^2$, where $x = \nu/0.4\ghz$, and $S_\nu$
is in mJy.  The distribution, $p(\alpha)$, of measured spectral slopes
is roughly Gaussian over $\alpha = 0$--4.0, with a peak at $\alpha =
1.5$--2 \citep{lorimer95,maron00}.  Utilizing the above pulsar
statistics, along with the steady-state assumption of the last
section, we now proceed to estimate the detectable fraction of pulsars
near Sgr A$^*$; our approach is quite similar to that of
\citet{cordes97}.

At high frequencies, the detection of shallow-spectrum pulsars is
favored \citep[e.g.,][]{johnston92,wex96}. It is encouraging that
$\simeq$10\% of pulsars with a measured spectrum have $\alpha = 0$--1,
although $p(\alpha < 1)$ is poorly constrained.  We can crudely
estimate the fraction, $f(>\!\! S_{\rm min})$, with flux densities
greater than $S_{\rm min}$ by restricting to the range $\alpha = 0$--1
and assuming a flat distribution, $p(\alpha < 1) = 0.1$.  After
evaluating a simple integral, we find that
\begin{equation}
f(>\!\! S_{\rm min}) \simeq 5\%\,
\left(\frac{S_{\rm min}}{10\mujy}\right)^{-1}
\left(\frac{\ln x}{\ln 25}\right)^{-1}
\end{equation}
when $x\gg 1$.  If $p(\alpha < 1)$ is taken to be a linear or
quadratic function that vanishes at $\alpha = 0$, each of which is
consistent with the data, then $f(>\!\! S_{\rm min}) \simeq 2$--3\% at
10\,GHz, and scales with frequency approximately as $(\ln
x)^{-(n+1)}$, where $n = 1$ (linear) or $n = 2$ (quadratic).  If
$\simeq$20\% of pulsars are beamed toward the Earth
\citep[e.g.,][]{lyne88}, we conclude that $\sim$1\% of the active
pulsars near Sgr A$^*$, or as many as 1--10 (see \S~\ref{sec:pop}),
may be detectable with current technology. These pulsars would most
likely have been born in the past several million years.  The planned
Square-Kilometer Array\footnote{\url{http://www.skatelescope.org/}}
(SKA) will easily achieve sensitivities of $\la$$1\,\mu{\rm Jy}$, and
could perhaps ultimately find $\ga$100 orbiting pulsars.


\begin{deluxetable*}{p{3.3cm}ccc}
\tabletypesize{\footnotesize}
\tablecolumns{4}
\tablewidth{0pt}
\tablecaption{Pulse Arrival-Time Delays
\label{tab:timing}}
\tablehead{
\colhead{Delay\tablenotemark{a}} &
\colhead{Amplitude} &
\colhead{Width} &
\colhead{References} 
}
\startdata
Roemer\tablenotemark{b}\dotfill & $\sim$$1\,{\rm day}\,\calm^{1/3}\calpo^{2/3}\sin i$ & $\sim$$1\yr\,\calpo$ & 1  \\[2mm]
Einstein\tablenotemark{c}\dotfill & $\sim$$1\,{\rm hr}\,\calm^{2/3} \calpo^{1/3}\,e$ & $\sim$$1\yr\,\calpo$ & 1  \\[2mm]
First-Order Shapiro\tablenotemark{d}\dotfill & $\sim$$30\s\,\calm |\ln[(1-e)(1 -\sin i)]|$ & $\sim$$1\yr\,\calpo (1-e)^{3/2}(\cos i)^{1/2}$ & 1  \\[2mm]
Second-Order Shapiro\tablenotemark{e}\dotfill & $\sim$$0.1\s\,\calm^{5/3}\calpo^{-2/3}(1-e)^{-1}/\cos i$ & $\sim$$1\yr\,\calpo (1-e)^{3/2}\cos i$ & 2, 3\\[2mm]
Frame Dragging\tablenotemark{f}\dotfill & $\sim$$0.1\s\,\calm^{5/3}\calpo^{-2/3}(1-e)^{-1}\chi/\cos i$ & $\sim$$1\yr\,\calpo (1-e)^{3/2}\chi \cos i$ & 2, 3, 4, 5\hspace{-12pt}\smallskip
\enddata

\tablenotetext{a}{The dimensionless variables used are $M_{6.5}
= M_{\rm BH}/10^{6.5}\msun$ and $P_1 = P_{\rm orb}/1\yr$. For
simplicity, we have adopted $\omega = 90^\circ$ in estimating the
amplitudes and widths.}
\tablenotetext{b}{\,\!Light travel time across the orbit.  The Keplerian orbit is evident in the Roemer delay.}
\tablenotetext{c}{Combined effect of time dilation and the gravitational redshift.}
\tablenotetext{d}{\,\!Lowest-order relativistic propagation delay in the gravitational
 field of a point mass.}
\tablenotetext{e}{Next highest order contribution to the propagation delay that is
independent of the BH spin.}
\tablenotetext{f}{Contribution to the net propagation delay due to the BH spin, in the special case 
where the spin direction is parallel to the orbital angular momentum of the pulsar.  Here 
$0 < \chi < 1$ is the dimensionless spin parameter, where $\chi = 1$ 
corresponds to an extreme Kerr BH.}

\tablerefs{
(1)~Damour \& Taylor 1992; 
(2)~Dymnikova 1986;
(3)~Goicoechea et al. 1992;
(4)~Laguna \& Wolszczan 1997;
(5)~Wex \& Kopeikin 1999
}

\end{deluxetable*}


\section{PULSAR TIMING AND RELATIVITY}\label{sec:timing}

The dynamics of a pulsar orbiting Sgr A$^*$ are revealed through
analysis of the pulse arrival times.  The Newtonian dynamical
signature of the supermassive BH should be clearly evident in a small
segment of the orbit, via the acceleration of the pulsar.  After a
full orbit, a best-fit Keplerian timing model is obtained.  Residuals
of the Keplerian solution contain further dynamical information,
including contributions from relativistic gravity.  In this section,
we quantify various arrival-time delays and secular effects, and
crudely assess their measurability.

\bigskip

\subsection{Time Delays and Secular Effects}\label{sec:delays}

Each pulse arrival time, $t$, at the solar-system barycenter is
related to the pulsar proper time, $t'$, by $t - t_0 = t' + \sum_i
\Delta_i$, where $t_0$ is a reference time, and the $\Delta_i$ are
variable delays due to the Keplerian motion and relativity.  The
delays are functions of $t'$ and the following Keplerian parameters:
(1) BH mass, $M_{\rm BH}$, (2) orbital period, $P_{\rm orb}$, (3)
eccentricity, $e$, (4) inclination, $i$, and (5) longitude of
pericenter, $\omega$.  When frame-dragging is considered, we must also
specify the magnitude and direction of the BH spin.  Each delay may be
characterized by an {\em amplitude}, the difference between the
maximum and minimum values of $\Delta_i$, and a {\em width}, the
timescale over which the delay shows the largest variation.

In Table \ref{tab:timing}, we quantify the five delays with the
largest expected amplitudes in the special case where $\omega =
90^\circ$, so that superior conjunction---when the pulsar is farthest
behind the BH---coincides with pericenter passage.  This simplifies
the analysis and illustrates roughly the dependence on eccentricity.
Dimensionless variables used in Table 1 are $\calm = M_{\rm
BH}/10^{6.5}\msun$ and $\calpo = P_{\rm orb}/1\yr$.  Not included in
Table~1 are the delays due to aberration of the beamed pulsar emission
\citep[e.g.,][]{smarr76} or the bending of light rays in the
gravitational field of the BH \citep{doroshenko95,wex99}.  Each of
these delays has an amplitude of $\la$1\,ms for a wide range of
$P_{\rm orb}$, $e$, and $i$.

Several secular processes cause changes in the orbital elements over
long timescales; these include the following.  Emission of
gravitational radiation by an orbiting pulsar causes the semimajor
axis to shrink on a timescale of
$\sim$$10^{13}\yr\,\calm^{-2/3}\calpo^{8/3}(1-e^2)^{7/2}$
\citep[e.g.,][]{taylor89}, which is typically too long to be of
interest.  The geodetic precession rate of the pulsar spin axis is
$\simeq$$0\fdg 1\,\calm^{2/3}\calpo^{-2/3}(1-e^2)^{-1}$ per orbit
\citep[e.g.,][]{weisberg89}.  This is evident as a long-term change in
the pulse profile, and its measurement depends on, among other things,
the precise geometry of the pulsar beam.  For $e \ga 0.9$ and $P_1
\sim 1$ it is conceivable that geodetic precession could be detected
after several orbits.  We now estimate the Newtonian and relativistic
contributions to the secular apsidal precession of the pulsar orbit.

Suppose that surrounding the BH is a spherically symmetric
distribution of matter, in the form of mostly low-mass stars and
compact objects (e.g., \S~\ref{sec:pert}).  For a density profile
$\rho \propto r^{-\gamma}$, the extended mass enclosed within a radius
$r$ is $M_e(r) \propto r^{3-\gamma}$ ($\gamma < 3$).  If
$M_e(a)/M_{\rm BH} \ll 1$ for an orbit with semimajor axis $a$, the
Newtonian contribution to the change in $\omega$ per orbit
is\footnote{We determined the precession rate by computing the
orbit-averaged rate of change of the Laplace-Runge-Lenz vector
\citep[e.g.,][]{goldstein80}, $\mathbf{e} = (GM_{\rm
BH})^{-1}\mathbf{v}\times \mathbf{h} - \mathbf{r}/r$, where
$\mathbf{h} = \mathbf{r}\times\mathbf{v}$.  \citet{weinberg72} uses
the same technique to calculate the general relativistic precession
rate.}
\begin{equation}
\Delta\omega_{\rm N} = \frac{M_e(a)}{M_{\rm BH}}
\frac{(1 - e^2)^{3-\gamma}}{e}
\int_0^{2\pi} d\phi \frac{\cos\phi}{(1+e\cos\phi)^{3-\gamma}}~,
\end{equation}
where $\phi$ is the true anomaly for the unperturbed elliptical
trajectory.  If $\gamma = 2$ (a plausible choice) the above integral
is analytic, and we have
\begin{equation}
\Delta\omega_{\rm N} = -2\pi 
\frac{M_e(a)}{M_{\rm BH}}
\frac{1-e^2}{e^2}\left[\frac{1}{(1-e^2)^{1/2}} - 1\right]~,
\end{equation}
where the minus sign indicates retrograde precession, which is
generally the case for $\gamma < 3$.  For example, if $\gamma = 2$,
$M_e(a)/M_{\rm BH} = 0.01$, and $e = 0.9$, we find that
$\Delta\omega_{\rm N} \simeq 1^\circ$.

The net apsidal precession rate also includes two relativistic
contributions.  In the Schwarzschild spacetime, the prograde advance
per orbit is \citep[e.g.,][]{weinberg72}
\begin{equation}
\Delta \omega_{\rm S} \simeq +0\fdg
23\,\calm^{2/3}\calpo^{-2/3}(1-e^2)^{-1}~.
\end{equation}
For a spinning BH, frame dragging introduces an additional
contribution \citep[e.g.,][]{jaroszynski98b,wex99}:
\begin{equation}
\Delta \omega_{\rm FD} \simeq -27\arcsec\,\calm
\calpo^{-1}(1-e^2)^{-3/2}\chi \cos\psi~,
\end{equation}
where $\psi$ is the angle between the angular momentum vectors of the
BH and the orbit, and $0<\chi < 1$ is the dimensionless BH spin
parameter.

\subsection{Remarks on Measurability}

The degree to which different contributions to the timing residuals
can be resolved depends on the precision and number of measured
arrival times.  Typical precisions are $\delta t \sim
(10^{-3}$--$10^{-2})P_p$, or $\sim$1--10\,ms for $P_p\simeq 1\s$.  If
one average arrival time is measured each day the pulsar is observed,
and $N \sim 100$ arrival times are measured per orbit ($P_{\rm orb}
\ga 1\yr$), then we expect a root-mean-square (RMS) timing precision
of $\epsilon \sim \delta t/\sqrt{N} \la 1$\,ms over one orbital
period.  However, various potential sources of error may limit the net
timing precision in a single orbit to $\sim$$P_p$.  In particular,
stochastic ``timing noise,'' which is largest for the youngest, most
luminous pulsars, may introduce net residuals of $\ga$$0.1P_p$ after
several years \citep[e.g.,][]{arzoumanian94}.  In general, a precise
assessment of the measurability of arrival-time delays and secular
effects requires detailed simulations that cover a large parameter
space.  Here we present some simple statements regarding the detection
of the delays listed in Table~1, as well as apsidal precession.

The amplitudes of the Einstein and first-order Shapiro delays given in
Table~1 suggest that these effects should be easily measurable over a
wide range in orbital parameters.  This may indeed be the case for the
Shapiro delay, allowing for an independent determination of $\sin i$,
as long as perturbations to the orbit by stellar encounters (see
\S~\ref{sec:pert}) do not have a significant impact.  However, the
Einstein delay can only be measured if the orbit undergoes appreciable
apsidal precession \citep[e.g.,][]{blandford76,damour92}. In effect,
extraction of the Einstein delay requires that the change in
$\cos\omega$ be resolved sufficiently, which is generally more
difficult than measuring $\dot{\omega}$ alone (see below).  For a
characteristic RMS timing precision of $\epsilon \la P_p$, the
second-order Shapiro and frame dragging delays may be just at the
threshold of detectability for $P_1 \simeq 1$, unless the orbit is
highly inclined and eccentric.

The rough scaling of the fractional error in the measured value of
$\dot{\omega}$ is \citep[e.g.,][]{blandford76}
\begin{equation}
\left|\frac{\delta\dot{\omega}}{\dot{\omega}}\right| 
\sim 10^{-3} \frac{k\epsilon}{N_{\rm orb} e \Delta\omega_d \sin i}
M_{6.5}^{-1/3} P_1^{-2/3}~,
\end{equation}
where $k$ is a dimensionless factor that depends on the initial values
of the Keplerian parameters, $N_{\rm orb}$ is the number of orbits
over which the pulsar is monitored, $\Delta\omega_d$ is the net
apsidal advance per orbit in degrees, and $\epsilon$ is in seconds.
Even under rather unfavorable circumstances, where, e.g., $k \sim
100$, it might be possible to obtain a $\la$10\% measurement of
$\dot{\omega}$ in only two or three orbits if $\Delta\omega_d \sim 1$,
thus facilitating the detection of the Einstein delay.  However, the
various contributions to $\dot{\omega}$ cannot be determined
independently.  The degeneracy could be broken if two or more pulsars
are monitored over several orbits; such an ambitious project probably
must wait for the SKA.


\section{DISCUSSION}\label{sec:dis}

Here we address three additional topics pertaining to observations of
a radio pulsar orbiting Sgr A$^*$.  We first discuss how pulsar timing
can be used to probe the physics of the accretion flow onto the BH.
This is followed by a short investigation of the effects of
gravitational interactions between an orbiting pulsar and the
surrounding cluster of stars and remnants.  Finally, we consider the
prospects of astrometrically monitoring an orbiting pulsar.

\bigskip

\subsection{The Interstellar Plasma Around Sgr A$^*$}\label{sec:plasma}

Timing observations of a pulsar orbiting Sgr A$^*$ can be used to
derive the properties of the local interstellar plasma.  As the pulsar
moves, photons trace many different lines of sight through the plasma
to the observer.  A gradient in the free-electron density on the scale
of the pulsar orbit introduces {\em dispersive} and {\em refractive}
pulse arrival-time delays that vary over the orbital
period\footnote{Here we are considering the gradient in the spatially
averaged electron density near Sgr A$^*$.  This is distinct from the
small-scale turbulent density fluctuations that are responsible for
angular and pulse broadening (see \S~\ref{sec:det}).}.  Each of these
effects has a characteristic frequency dependence, pointing to a need
for multi-frequency observations.  Because the detection of
shallow-spectrum pulsars is favored at high frequencies (see
\S~\ref{sec:det}), there is a good chance that a pulsar detected at
10\,GHz will also be detected at 15--20\,GHz for the same sensitivity.

{\em Chandra} observations of the Galactic center \citep{baganoff03}
indicate that the density and temperature of the electrons at
$\simeq$$1\arcsec$ ($\simeq$8000\,AU) from Sgr A$^*$ are $n_e \sim
100\cm^{-3}$ and $kT_e \simeq 1$--2\,keV.  The gravitational potential
of the BH exceeds the thermal energy of the plasma inside the
\citet{bondi52} radius, $R_{\rm B} \sim GM_{\rm BH}/c_s^2$, where $c_s
\sim (kT_e/m_p)^{1/2}$ is the thermal speed, assuming equipartition
between electrons and protons, where $m_p$ is the proton mass.  For
the plasma around Sgr A$^*$, we find that $R_{\rm B} \sim 10^4\au$.
Within $R_{\rm B}$, $n_e(r)$ depends on the physics of the accretion
flow.  We adopt $n_e(r) \sim 10^2\cm^{-3} (10^4\au/r)^{\beta}$, where
models predict $\beta \simeq 1$--1.5 \citep[e.g.,][]{melia01,yuan03}.

The plasma frequency is $\nu_p = (n_e e^2/\pi m_e)^{1/2}\simeq
90\,{\rm kHz}\,(10^4\au/r)^{\beta/2}$ for the density profile given
above.  The index of refraction of the plasma is $\xi(\nu) \simeq 1 -
(\nu_p/\nu)^2/2$, for $\nu \gg \nu_p$.  Radiation at different
frequencies propagates through the medium with different group
velocities of $c\xi(\nu)$, so that the arrival time of a pulse is
frequency dependent.  The difference in arrival times at frequencies
$\nu_1$ and $\nu_2$ is proportional to $(\nu_1^{-2} - \nu_2^{-2})$DM,
where DM is the {\em dispersion measure}, the column density of free
electrons along the path of the pulse.  Modulation of DM over the
orbital period of a pulsar bound to Sgr A$^*$ can thus be used to
constrain the density profile of the plasma within the orbit.

Refraction due to the large-scale gradient of the electron density
causes a net angular deflection of individual pulses that reach the
observer.  Consequently, there will be a variable {\em geometrical}
time delay as the pulsar orbits.  We assume that a light ray is
refracted impulsively as it passes its closest approach, $b$, to Sgr
A$^*$---the ``scattering-screen'' or ``thin-lens'' approximation.  The
deflection angle is $\theta_p \sim \beta[\nu_p(r = b)/\nu]^2/2$,
pointing {\em away} from Sgr A$^*$.  From the thin-lens geometry, the
resulting excess propagation time, compared to that of a straight-line
path, is $\sim$$r\theta_p^2/2c\propto \nu^{-4}$, where $r$ is the
orbital radius.  For an orbital period of 10\,yr, inclination of $i =
80^\circ$, and eccentricity of $e = 0.9$, the minimum possible impact
parameter is $a(1-e)\cos i \simeq 10\au$, where $\nu_p(10\au) \simeq
16\,{\rm MHz}$ for $\beta = 3/2$.  At $\nu = 10\ghz$, the amplitude
and width of the refractive delay are, respectively, $\sim$0.1\,s and
$\sim$10 days.  The strong frequency dependence of the refractive
delay distinguishes it from dynamical effects.

\subsection{Dynamics of the Sgr A$^*$ Cluster}\label{sec:pert}

A given orbiting pulsar will interact gravitationally with an unknown
number of normal stars, other NSs, white dwarfs, and stellar-mass
BHs. An important prediction of our work is that thousands of NSs born
over the past $\sim$1\,Gyr may orbit Sgr A$^*$ with periods of
$\la$100\,yr.  In addition, \citet{miralda00} suggest that
$\sim$$10^4$ stellar-mass ($\simeq$$10\msun$) BHs may have migrated
due to dynamical friction to the central $\simeq$1\,pc about Sgr
A$^*$; extending their results, we find that perhaps $\sim$$10^2$ such
BHs may reside in the central $1\arcsec$.  \citet{chaname02} have
discussed the possibility of using positional information for $\sim$50
{\em millisecond} radio pulsars within a few parsecs of Sgr A$^*$ to
indirectly infer the presence of a large population of stellar-mass
BHs.

For a star orbiting Sgr A$^*$, many weak gravitational perturbations
accumulate over a relaxation time to yield a significant net change in
the orbital parameters\footnote{This is due to the granularity of the
stellar distribution. Long-term orbital precession, as discussed in
\S~\ref{sec:delays}, is due to the smoothed potential of the extended
stellar mass component.}.  In the absence of resonant effects, it is
straightforward to show that near Sgr A$^*$ the relaxation time is
$\tau_{\rm rel} \sim (10 N_s)^{-1}\,(M_{\rm BH}/M_s)^2 P_{\rm orb}$,
where $M_s$ and $N_s$ are the the typical mass and total number,
respectively, of the perturbing stars \citep[e.g.,][]{rauch96}.  If
$M_{\rm BH}/M_s \simeq 10^6$ and $N_s = 10^3$--$10^4$, we find that
$\tau_{\rm rel} \sim (10^7$--$10^8) P_{\rm orb}$.  Resonant
angular-momentum relaxation \citep{rauch96}, which causes variation in
only the orbital eccentricity and orientation angles, can act on a
much shorter timescale of $\tau_{\rm rel, res} \sim (M_{\rm
BH}/M_s)P_{\rm orb} \sim 10^6P_{\rm orb}$.

Random-walk fluctuations in the orbital energy and angular momentum
lead to a root-mean-square {\em fractional} change in one orbit of
approximately $(P_{\rm orb}/\tau)^{1/2}\sim 10^{-3}$--$10^{-4}$, where
$\tau$ is either the conventional or resonant relaxation time. The
resulting change in $a/c$ may be $\sim$100--1000 light seconds for
$P_{\rm orb}\sim 10\yr$.  Gravitational perturbations by stars and
remnants may have measurable consequences for the timing analysis of a
pulsar orbiting Sgr A$^*$, possibly even masking the important
relativistic time delays discussed in the last section.  With the SKA
it may be possible to obtain a large sample of regularly timed,
orbiting radio pulsars.  If evidence of random gravitational
encounters is found in the timing properties of a significant fraction
of these pulsars, important clues could be extracted regarding the
number, masses, and velocities of the Sgr A$^*$ cluster members.

\subsection{Radio Astrometry}

It would be an extreme challenge to image a pulsar orbiting Sgr A$^*$
with an interestingly short orbital period.  The projected apocenter
separation is less than $a(1+e)/D = 20\,{\rm
mas}\,\calm^{1/3}\calpo^{2/3}(1+e)$.  Very long baseline
interferometry (VLBI) would be required to resolve a faint pulsar next
to the very bright Sgr A$^*$ \citep[$\sim$1\,Jy from 1 to
10\,GHz;][]{melia01}.  Such observations would have to be conducted at
$\nu \ga 10\ghz$ to sufficiently reduce the scattering diameter of Sgr
A$^*$ (see \S~\ref{sec:det}).  With a current VLBI sensitivity of
$\simeq$1\,mJy, probably no orbiting pulsars would be detectable (see
\S~\ref{sec:det}).  The SKA will have the requisite sensitivity, and
the current design specifications call for an angular resolution of
$\simeq$10\,mas at 10\,GHz, and, of course, higher astrometric
resolution.  Astrometric precisions smaller than $\simeq$0.1\,mas at
$\sim$10\,GHz may not be attainable from the surface of the Earth
\citep[e.g.,][]{chatterjee04}.  The advent of highly sensitive space-
or Moon-based VLBI instruments is then a necessary step toward
$\la$$10\mu{\rm as}$ astrometry for Sgr A$^*$ pulsars.

If $\Delta\omega_d$ is the net apsidal advance per orbit in degrees,
then the angular shift of the apocenter position in one orbit is
\begin{equation} 
\Delta\theta_a < 0.34\,{\rm mas}\,\Delta\omega_d M_{6.5} 
P_1^{2/3}(1+e)~.  
\end{equation}
For a large eccentricity and $P_1 \sim 10$, $\Delta\theta_a$ could be
$\sim$1\,mas.  It is at least conceivable that apsidal precession
could be detected from the ground after several orbits.

If the orbit is inclined with respect to the BH spin, frame dragging
causes the line of nodes, $\Omega$, on the plane of the sky to advance
in one orbit by an amount \citep[e.g.,][]{jaroszynski98b}
\begin{equation}
\Delta \Omega_{\rm FD} \simeq 9\arcsec\,\calm 
\calpo^{-1}(1-e^2)^{-3/2}\chi \sin\psi~,
\end{equation}
where $\chi\sin\psi$ is the projection of the spin onto the orbital
plane.  The corresponding apocenter shift is expected to be no larger
than a few tens of $\mu{\rm as}$.  If the contributions from frame
dragging to apsidal (see \S~\ref{sec:timing}) and nodal precession are
measured, it would be possible to determine the magnitude and
direction of the BH spin, but this is a highly unlikely prospect.

The effects of gravitational lensing by the BH may be important when
the pulsar is near superior conjunction.  For lensing by a point mass,
two images are produced, one inside and one outside the Einstein
radius, $\theta_{\rm E} \simeq (4GM_{\rm BH}r/c^2D^2)^{1/2}$, where $r
\ll D$ is the orbital radius near superior conjunction.  For $r =
100$--1000\,AU, we find $\theta_{\rm E} \simeq 0.5$--$1.5\,{\rm mas}$.
Since $\theta_{\rm E}$ is small, we consider here only the image
outside $\theta_{\rm E}$.  At superior conjunction, the angular
separation between the pulsar and its image is $\delta\theta\simeq
\theta_{\rm E}^2D/b$, where $b = r\cos i$ is the impact parameter
\citep[e.g.,][]{jaroszynski98a}.  We then find that $\delta\theta\sim
20\,\mu{\rm as}/\cos i$, which is $\sim$1\,mas for $i = 89^\circ$.
Therefore, gravitational lensing has a small astrometric signature in
typical situations.


\acknowledgements

We are grateful to D. Chakrabarty, B. Gaensler, S. Hughes, S. Ransom,
S. Rappaport, and M. Reid for useful discussions, and F. Camilo and
V. Kaspi for reading an early draft of the paper and providing
valuable comments.  We also thank the anonymous referee for providing
comments that led to significant improvements of the paper.  EP was
supported by NASA and the Chandra Postdoctoral Fellowship program
through grant number PF2-30024.  AL acknowledges support from the John
Simon Guggenheim Memorial Fellowship, as well as NSF grants
AST-0071019 and AST-0204514, and NASA grant NAG5-13292.




\end{document}